\documentclass{aastex}
\usepackage{emulateapj5}
\usepackage{onecolfloat5}
\usepackage{apjfonts}

\def\myputfigure#1#2#3#4#5#6%
{\vskip#5pt\makebox[0pt]{\hskip#2in
\includegraphics[width=#3\textwidth,angle=#6]{#1}}\vskip#4pt\hfill}

\def\gsim{\;\rlap{\lower 2.5pt
 \hbox{$\sim$}}\raise 1.5pt\hbox{$>$}\;}
\def\lsim{\;\rlap{\lower 2.5pt
   \hbox{$\sim$}}\raise 1.5pt\hbox{$<$}\;}

\newcommand{\nf}{x_{\rm \scriptscriptstyle H}}
\newcommand{\ionf}{x_{\rm \scriptscriptstyle i}}
\def\la{\lower.5ex\hbox{$\; \buildrel < \over \sim \;$}}
\def\ga{\lower.5ex\hbox{$\; \buildrel > \over \sim \;$}}
\def\apj{ApJ}

\def\apjl{ApJL}

\def\aj{AJ}
\def\aap{A \& A}

\def\physrep{Physics Reports}

\begin{document}
\twocolumn[
\title{Can We Detect the Anisotropic Shapes of Quasar HII Regions During Reionization\\ Through The Small--Scale Redshifted 21cm Power Spectrum?}

\author{Shiv Sethi$^{1}$ \& Zolt\'{a}n Haiman$^{2}$}
\affil{
$^1$Raman Research Institute, C. V. Raman Avenue, Sadashivanagar, Bengalooru, 560 080, India; sethi@rri.res.in\\
$^2$Department of Astronomy, Columbia University, 550 West 120th Street, New York, NY 10027, USA; zoltan@astro.columbia.edu
}
\vspace{+0.4cm}

\begin{abstract}
Light travel time delays distort the apparent shapes of HII regions
surrounding bright quasars during early stages of cosmic reionization.
Individual HII regions may remain undetectable in forthcoming
redshifted 21 cm experiments. However, the systematic deformation
along the line of sight may be detectable statistically, either by
stacking tomographic 21cm images of quasars identified, for example,
by {\it JWST}, or as small--scale anisotropy in the three--dimensional
21cm power spectrum.  Here we consider the detectability of this
effect.  The anisotropy is largest when HII regions are large and
expand rapidly, and we find that if bright quasars contributed to the
early stages of reionization, then they can produce significant
anisotropy, on scales comparable to the typical sizes of HII regions
of the bright quasars ($\lsim 30 \, \rm Mpc$).  The effect therefore
cannot be ignored when analyzing future 21cm power spectra on small
scales.  If $10$ percent of the volume of the IGM at $z\simeq 10$ is
ionized by quasars with typical ionizing luminosity of $S\gsim 5\times
10^{56} \, \rm sec^{-1}$, the distortions cause a $\gsim 10$ percent
enhancement of the 21cm power spectrum in the radial (redshift)
direction, relative to the transverse directions. The level of this
anisotropy exceeds that due to redshift--space distortion, and has the
opposite sign.  We show that on-going experiments such as MWA should
be able to detect this effect.  A detection would reveal the presence
of bright quasars, and shed light on the ionizing yield and age of the
ionizing sources, and the distribution and small--scale clumping of
neutral intergalactic gas in their vicinity.
\end{abstract}
\vspace{\baselineskip}
]

\section{Introduction}

How and when the intergalactic medium (IGM) was reionized is one of
the long outstanding questions in cosmology, likely holding many clues
about the nature of the first generation of light sources and the end
of the cosmological ``Dark Age'' (see, e.g., Barkana \& Loeb 2001 for
a review).  

Observational break-throughs in recent years have revealed our first
clues about this epoch.  The Thomson scattering optical depth, $\tau_e
= 0.09\pm0.03$ (Page et al. 2006; Spergel et al. 2006), inferred from
the polarization anisotropies in the cosmic microwave background (CMB)
by the {\it Wilkinson Microwave Anisotropy Probe} ({\it WMAP}),
suggests that reionization began at redshift $z\sim10$.  The detection
of the so--called Gunn-Peterson (GP) troughs in the spectra of
high-redshift quasars discovered in the Sloan Digital Sky Survey
(White et al. 2003; Fan et al. 2006), and more detailed analyses of
these spectra (Mesinger \& Haiman 2004; 2006), suggest that
reionization is ending at $z\sim6$.

Despite this progress, little is known about the nature of the first
ionizing sources.  The dearth of bright quasars at $z\sim 6$, and a
tight limit on the contribution of lower--luminosity quasars to the
unresolved soft X--ray background (Dijkstra, Haiman \& Loeb 2004)
requires that the ionizing background at $z\sim 6$ be dominated by
stars, rather than quasars.  On the other hand, luminous quasars may
still contribute up to $\sim 10\%$ of the ionizing background at
$z\approx 6$ (e.g. Sbrinovsky \& Wyithe 2007), and the contribution of
the ionizing radiation from accreting quasar black holes may have been
more significant at higher $z$, during the earliest stages of
reionization (Madau et al. 2004; Ricotti \& Ostriker 2004; Oh 2001;
Venkatesan, Giroux \& Shull 2001).  Overall, the present data allow a
wide range of possible reionization histories, driven by different
sources and various physical feedback mechanisms (e.g. Haiman \&
Holder 2003).

One of the promising future probes that could constrain, especially
the early stages, of reionization, is based on the detection of the
redshifted 21-cm line of neutral hydrogen (see a recent review by
Furlanetto, Oh \& Briggs 2006).  Forthcoming 21cm surveys are expected
to deliver tomographic maps of the brightness temperature of the 21cm
line, and detect fluctuations corresponding to the topology of large
discrete HII regions.  Numerous recent studies have addressed various
aspects of the power spectrum of such brightness temperature
fluctuations (see references in Furlanetto et al. 2006).

In this paper, we examine one particular effect, arising from the
distortion in the apparent shapes of HII regions due to the
light--travel time delay between a light source and the ionization
front it drives into the IGM.  Since the distance from the Earth to
various points on the surface of an HII region differ, different
patches of an expanding HII region will be observed at different
stages of their evolution, corresponding to different light travel
times. As pointed out in Cen \& Haiman (2000), the delay can be
ignored along the line of sight to individual sources (for example,
when computing Lyman $\alpha$ absorption spectrum of a point source;
see also White et al. 2003). On the other hand, the delay modifies the
apparent expansion of a resolved HII region in the transverse
direction, as could be observed, for example, in 21cm studies (Wyithe
\& Loeb 2004c).  The full equal--arrival time surface in three
dimensions has been computed in simple models of individual spherical
quasar HII regions by Yu (2005), and shown to be highly anisotropic
for young and bright sources.

In principle, the distortion could be measured directly in tomographic
images of individual HII regions (Wyithe, Loeb \& Barnes 2005),
although the immediate next generation of 21cm instruments (such as
PAST or MWA) are unlikely to be able to achieve the required S/N for
such direct imaging.  Since individual HII regions are likely to be
intrinsically highly anisotropic, such a measurement can only be
performed statistically.  In this paper, we examine and quantify the
possibility of measuring the corresponding anisotropy in the
three---dimensional power spectrum of the 21cm brightness temperature.
This method does not require that individual HII regions be detectable
at high S/N.

Barkana \& Loeb (2006) recently studied the anisotropy in the
three--dimensional 21cm power spectrum that arises from finite
light--travel time across larger, joint HII regions that have merged
around galaxies.  The growth of such super--bubbles is dictated by the
collapse rate of cosmic structures, and it produces an anisotropy
which will be significant on larger scales (comparable to the sizes of
the HII super--bubbles).  In comparison, the effect we describe here
arises from the finite rate of growth of isolated HII regions around
individual quasars.  Finite--speed--of light effects during the
lifetime of quasars can produce an appreciable additional
smaller--scale anisotropy only in a likely narrow window of redshifts
during reionization: on the one hand, bright quasars need to present,
which occurs only at relatively late stages, but on the other hand,
once the ionized bubbles around these bright quasars percolate
significantly, the small--scale anisotropy will diminish.  Conversely,
a detection of any small-scale anisotropy would reveal the presence of
bright quasars, and shed light on the ionizing yield and age of the
ionizing sources, and the distribution and small--scale clumping of
neutral intergalactic gas in their vicinity.  In this paper, we
quantify the conditions under which a detection of this additional
anisotropy, due to the presence of quasars, could be feasible with
forthcoming experiments, with specifications similar to that proposed
for MWA.

The rest of this paper is organized as follows.
In \S~\ref{sec:fluctuations}, we discuss the basic formalism to
compute the 21cm brightness temperature fluctuations, including the
anisotropy from the light travel time delay.
In \S~\ref{sec:results}, we present the expected level of anisotropy
in a few simple models, and discuss how it is expected to vary with
various parameters.
In \S~\ref{sec:noise}, we discuss instrumental noise and the
corresponding limits on the detectability of the anisotropy.
In \S~\ref{sec:discussion}, we discuss our results and the
implications of this work.
Finally, in \S~\ref{sec:conclude}, we summarize our main conclusions.
Throughout this paper, we adopt the background cosmological parameters
$\Omega_m=0.29$, $\Omega_{\Lambda}=0.71$, $\Omega_b=0.047$,
$H_0=72~{\rm km~s^{-1}~Mpc^{-1}}$, with a power spectrum normalization
$\sigma_{8h^{-1}}=0.8$ and slope $n_s=1$, consistent with the values
measured recently by the {\it WMAP} experiment (Spergel et al. 2007).
Unless noted otherwise, all lengths below are quoted in comoving
units.

\section{Brightness Temperature Fluctuations}
\label{sec:fluctuations}

Forthcoming radio observations will measure the brightness temperature
at the redshifted wavelength $(1+z)21$cm.  The brightness temperature
is expected to be below that of the CMB before the first luminous
sources turn on, so that cosmic HI can only be seen in absorption.
Once the first sources turn on and ionize and heat the IGM, the
brightness temperature can rise above that of the CMB, so that the
21cm signal is seen in emission (see Furlanetto et al. 2006 for a
review and references).  Here we are interested in the epoch of
reionization, when the spin temperature $T_s \gg T_{\rm CMBR}$.  In
this limit, the brightness temperature $T$ of the redshifted HI line
is independent of $T_s$, and is fully determined by the local density
of neutral HI (see e.g. Field 1959; Madau, Meiksin \& Rees 1997; Sethi
2005).  In this section, we outline how we compute the anisotropic
correlation function of the spatially varying brightness temperature.

\subsection{Basic Formalism for Temperature Fluctuations}

Observations will measure the brightness temperature $T$ at a given
frequency $\nu$ and direction on the sky $\hat n$, but in a narrow
frequency band, the signal is dominated by the emitting gas at the
corresponding redshift $z$, or radial distance $r$ (with the relation
between $r$ and observed frequency $\nu$ obtained from the appropriate
cosmological model). Given the telescope beam $B(\hat n, \hat
n^\prime)$, which for radio interferometers will correspond to the
synthesized beam, we obtain:
\begin{equation}
T(\hat n,r) = \int d\Omega^\prime \int dr^\prime T_0(r^\prime) W_r(r^\prime) \psi(\hat n^\prime,r^\prime) B(\hat n, \hat n^\prime),
\end{equation}
where $T_0(r)$ is the mean brightness temperature at the redshift
$z=z(r)$,
\begin{eqnarray}
T_0 &=& {3 \over 32 \pi}{h \lambda^2 A_{21} n_{\rm b}(z) \over k H(z)}
  \approx \\ 
& \approx & 40 \left [\left({0.127 \over \Omega_m h^2} \right ) \left
  ({\Omega_b h^2 \over 0.223} \right)^2 \left ({ 0.73 \over h} \right
  )^2 \left ({1+z \over 11}\right ) \right ]^{1/2}{\, \rm mK},
  \nonumber
\end{eqnarray}
and 
\begin{equation}
 \psi = \nf({\bf r}) [1+ \delta({\bf r})], 
\end{equation}
with $\delta$ and $\nf$ denoting the spatially fluctuating
over-density and neutral hydrogen (HI) fraction ($n_{\rm
\scriptscriptstyle HI}/[n_{\rm \scriptscriptstyle HI} +n_{\rm
\scriptscriptstyle HII}]$), respectively.  Finally, $W_r(r^\prime)$ is
a window function, peaking at $r$, and representing the bandwidth of
the observations.

Assuming for simplicity that $T_0(r)$ does not evolve significantly
within the bandwidth, it can be taken out of the integral. In this case,
\begin{equation}
T(\hat n,r) =T_0(r)\int d\Omega^\prime \int dr^\prime  W_r(r^\prime) \psi(\hat n^\prime,r^\prime) B(\hat n, \hat n^\prime),
\end{equation}
where $r$ corresponds (for example) to the center of the observed
frequency band. The two--point correlation of the temperature
fluctuation can then be written as
\begin{eqnarray}
\nonumber
&\langle \Delta T(\hat n_1,r_1) \Delta T(\hat n_2,r_2) \rangle = T_0(r_1)T_0(r_2)\times\\
&\,\nonumber \times\int dr \int dr' \int d\Omega d\Omega^\prime W_{r_1}(r) W_{r_2}(r') B(\hat n,\hat n_1)B(\hat n^\prime,\hat n_2)\\
&\, [\langle \psi(\hat n_1,r) \psi((\hat n_2,r') \rangle - \langle \psi \rangle^2].
\end{eqnarray}
Note that statistical homogeneity of the signal is assumed but not
statistical isotropy, i.e. the correlation depends only on
$\bf{r_1-r_2}$, and the brackets denote averages over (say)
$\bf{r_1}$. The expression above also assumes that the neutral
fraction does not evolve significantly over the length--scales of
interest, so $\langle \psi \rangle \equiv \langle \psi
\rangle_{\bf{r_1}}\approx \langle \psi \rangle_{\bf{r_2}}$.  It is
customary to assume a relatively broad bandwidth of $\simeq
0.2\hbox{--}0.5 \, \rm MHz$ while computing the theoretical signal
(e.g. Zaldarriaga et~al. 2004).  This choice is motivated by the fact
that it roughly matches the length scales ($\simeq 10 \, \rm Mpc$)
corresponding to the synthesized beam ($4'$) of the upcoming radio
interferometers, such as MWA.  However, the frequency resolution of
future experiments is much narrower, $\la 1 \, \rm kHz$ (by design,
this is necessary to remove radio frequency interference [RFI]).  This
corresponds to scales much smaller than other length scales in the
problem.  Mellema et al. (2006) have shown that the power spectrum on
the scales we will be interested in (i.e. on the characteristic size
$\gsim 10 \, \rm Mpc$, of the HII regions), is insensitive to the
frequency resolution once it is well below the above value.
Furthermore, for a statistical signal, increasing the bandwidth around
a given frequency does not give any advantage in the signal--to--noise
ratio for the detection (see discussion below).  As shown in
section~\ref{sec:results}, the scales of interest to us are generally
greater than $20 \, \rm Mpc$ and the effect of finite beam width of
$4'$ is unlikely to effect the signal significantly at those scales.
Therefore, for simplicity, and in accordance with our objectives in
this paper, we assume, $W_r(r^\prime) = \delta_{\scriptscriptstyle
D}(r^\prime -r)$ and $B(\hat n, \hat n^\prime) =
\delta_{\scriptscriptstyle D}(\hat n^\prime -\hat n)$.  We do note,
however, that the random placement of the resolution pixels relative
to the boundaries of the quasar bubbles will cause a shot noise in
estimates of the anisotropy that could become significant, and will
certainly necessitate averaging over many bubbles to detect angular
structure.  Quantifying this shot noise term is beyond the scope of
the present paper (averaging over a large number of bubbles will be
required, in any case, for other reasons; see the discussion in
\S~\ref{sec:discussion} below).

Under this approximation, the expression for the two--point correlation
function is simplified to
\begin{eqnarray}
\nonumber 
C(r_{12}, \theta) &\equiv& \langle \Delta T(\hat n_1,r_1)\Delta T(\hat
n_2,r_2) \rangle = \\
 & = & T_0(r_1)T_0(r_2) [\langle \psi(\hat n_1,r_1) \psi(\hat n_2,r_2) \rangle - \langle \psi \rangle^2].
\label{corr_fun_f} 
\end{eqnarray}
Here $\theta$ is the angle between the line of sight $\hat n$ and the
vector $\bf{r_1}-\bf{r_2}$ separating the two points, and brackets
denote average over (say) $\bf{r_1}$ (as noted above, the correlation
function could be written equivalently as a function of the angle
$\theta$ and the frequency difference $\nu_2-\nu_1$).  We caution the
reader that, as defined above, the angle $\theta$ is different from
the angle subtended on the sky. The latter is typically defined in
studies involving angular correlation functions, but we find the angle
relative to the line-of-sight, which directly follows the shapes of
the quasar--driven bubbles, more convenient for our
purposes.\footnote{We note that finite instrumental resolution may, in
practice, cause a degradation of the precision to which the anisotropy
of the correlation function can be extraced from the data, on scales
near the resolution, since the brightness temperature fluctuations
will effectively be smoothed with a filter that is defined in the
observed ($r,\hat{n}$) coordinates, and not in $(r,\theta)$
coordinates we use to specify the anisotropy.}

The correlation function of $\psi$ can be expanded as 
\begin{eqnarray}
\langle \psi({\bf r_1})\psi({\bf r_2}) \rangle &=& 
\langle 
\{ \nf({\bf r_1}) [1+ \delta({\bf r_1})]\}
\{ \nf({\bf r_2}) [1+ \delta({\bf r_2})]\}
\rangle \nonumber \\
& = & 
 \langle \nf({\bf r_1})  \nf ({\bf r_2}) \rangle + 
2\langle \nf({\bf r_1}) \delta({\bf r_1}) 
\nf({\bf r_2}) \rangle \nonumber \\ 
 && + \langle \nf({\bf r_1}) \delta({\bf
r_1})\nf({\bf r_2}) \delta({\bf r_2}) \rangle
\end{eqnarray}
The three and four-point functions in the expression above are, in
general, nonzero, and cannot be easily computed, as $\nf$ is not a
Gaussian random variable (for details, see, e.g., Furlanetto
et~al. 2004).

In what follows, we will drop the terms that contain
cross-correlations between the density and the neutral fraction (we
will discuss the neglect of these terms in detail below).  In this
case,
\begin{equation}
\langle \psi \psi \rangle - \langle \psi \rangle^2  =  
 \xi_{xx}
\xi_{\delta\delta}(r_{12}, \theta,z) +
\xi_{xx}  - \bar \nf ^2
\label{corr_fun_psi}
\end{equation}
Here $\xi_{\delta\delta} = \langle \delta({\bf r_1})\delta({\bf r_2})
\rangle$ is the two-point correlation function of the total density
contrast (which we compute using the fitting formula for the linear
spectrum in Eisenstein \& Hu 1999), and $\xi_{xx} = \langle \nf({\bf
r_1}) \nf({\bf r_2}) \rangle$, to be discussed in detail below, is the
correlation function owing to the inhomogeneities of the neutral
fraction.\footnote{ Another commonly used definition of the
correlation function is $\xi_{xx} = \langle \nf({\bf r_1}) \nf({\bf
r_2}) \rangle - \langle \nf \rangle^2$. We caution the reader that the
definition we adopt here is different.}  To make further progress, we
need to make some assumptions about the spatial variation of the
neutral fraction.

Before we proceed to compute the anisotropy in the above power
spectrum due to the non-spherical shapes of the individual HII
regions, we note that a different source of anisotropy will already be
produced by redshift space distortions due to peculiar velocities
(e.g. Bharadwaj \& Ali 2004; Barkana \& Loeb 2005). This effect is
analogous to the redshift--space distortion of the matter power
spectrum (Kaiser 1987), and is described in the linear regime by
\begin{eqnarray}
\nonumber
\xi_{\delta\delta}(r_{12},\theta,z) & \simeq & 
\xi_{\delta\delta}(r_{12},0,z) \left(1 + {2\over 3}\beta + {1\over 5}\beta^2 \right) + \\
&& + \xi_{\delta\delta}(r_{12},2,z) \left({4 \over 3} \beta + {4\over7} \beta^2 \right )P_2(\theta)
\label{corr_fun_rd}
\end{eqnarray}
Here $\beta \simeq \Omega_m^{0.6}/b$, we use $b=1$ throughout;
 $P_2(\theta)$ is the Legendre function with $\ell = 2$, and
\begin{eqnarray}
\xi_{\delta\delta}(r,0,z) & = & {D_{+}^2(z)\over 2\pi^2} \int dk k^2 P(k) j_0(kr) \\
\xi_{\delta\delta}(r,2,z) & = & -{D_{+}^2(z)\over 2\pi^2} \int dk k^2 P(k) j_2(kr). 
\end{eqnarray}
Here $D_{+}(z)$ is the growing mode of density perturbations and
$P(k)$ is the matter power spectrum.  We have further neglected the
fourth (and higher) moments of the correlation function, which make a
negligible contribution (see, e.g., Hamilton 1998).  The anisotropy we
consider below, intrinsic to the correlation function of the neutral
fraction, will be in addition to this redshift--space--distortion
anisotropy.

\subsection{Correlation Function of the Neutral Fraction}

While the fluctuations in the over-density $\delta({\bf r})$ are
specified by the cosmological initial conditions, in order to describe
the spatial variations in the neutral fraction $\nf({\bf r})$, we need
to make assumptions about the process of reionization. It is generally
believed that this transition occurred by the percolation of ionized
bubbles around individual sources, or clusters of sources, that formed
in nonlinear halos.  The nature of sources of ionization is not clear
and they could either be star-forming galaxies or QSOs, or both.
Several works have proposed simple models for the correlation function
$\xi_{xx}$ of the neutral fraction in this picture (e.g. Knox et
al. 1998; Gruzinov \& Hu 1998; Santos et al. 2003; FZH04; Zhang et
al. 2007).

We follow the simplest model described in FZH04 (adopted from Knox et
al. 1998), which assumes that the ionizing sources are randomly
distributed in space.  In reality, the sources are likely to be
located at the peaks of the density field and therefore clustered.  In
general, this will increase $\xi_{xx}$ (by about an order of
magnitude, for ionizing sources in dark matter halos that correspond
to 2--3$\sigma$ peaks; e.g. Santos et al. 2003), and can also modify
its anisotropy (see discussion in \S~\ref{sec:discussion} below).  In
the case of randomly distributed identical sources with space density
$n$, each surrounded by an ionized bubble with volume $V$, we have the
joint probability for the {\it ionized} fraction $\ionf=1-\nf$,
\begin{equation} 
 \langle \ionf({\bf r_1}) \ionf({\bf r_2}) \rangle =
(1- e^{-nV_o}) + e^{-nV_o}[1- e^{-n(V-V_o)}]^2,
\label{eq:overlap}
\end{equation}
where $V_o=V_o({\bf r_1-r_2})$ is the volume of the overlap between
two ionized bubbles located at ${\bf r_1}$ and ${\bf r_2}$.  The first
term in equation~(\ref{eq:overlap}) represents the probability that
both points are ionized by the same source; the second term describes
the case when they are ionized by two different sources.  Defining
$p_{\rm same}\equiv 1-\exp(-nV_o)$ as the probability that the two
points belong to the same ionized bubble (i.e., when the overlap
region contains at least one source), the correlation function for the
neutral fraction, obtained from the above equation, can be
conveniently re--written (Barkana \& Loeb 2006) as
\begin{equation} 
\nonumber
 \xi_{xx}  = 
 {\bar \nf({\bf r_1})\bar \nf({\bf r_2}) \over (1-p_{\rm same})}.
\end{equation}

We note that equation~(\ref{eq:overlap}) assumes that the bubble
boundaries are ``static'' and not affected by overlap.  A more
realistic assumption would be that the total ionized volume is
conserved, which would require expanding the joint boundaries of two
bubbles that have merged. This would guarantee $(1-\bar\nf)=nV$, but
modifying equation~(\ref{eq:overlap}) to take the bubble mergers into
account would require further assumptions and would be overly
complicated.  Alternatively, one could start with the ``postulate''
that bubbles do not overlap (i.e., counting a merged bubble as a
single object); however, this condition would not allow the bubbles to
be randomly distributed on small scales.

The above ambiguity affects the correlation function on scales
comparable to the bubble size, but we expect it would not significantly
modify its predicted anisotropy.  In any case, mergers between bubbles
will likely dilute the anisotropy signal we discuss below, and our
focus therefore will be at the earliest epochs of reionization, when
the filling factor of the ionized bubbles is small.  In this limit,
$nV\ll 1$, overlapping bubbles are rare, and the ambiguity is avoided:
we can approximate $p_{\rm same}\approx nV_o \equiv n fV \approx
(1-\bar\nf) f$, where $f\equiv Vo/V$.

If the ionized bubbles were spherically symmetric with $V=(4\pi/3)
R^3$ then we would have the simple expression
\begin{equation}
f(r,R) = 1 - {3 r \over 4 R^3}\left(R^2 - {1\over 12}r^2 \right),
\label{eq:frR}
\end{equation}
for $r \le 2R$ and zero otherwise, and where $r = \mid {\bf r_1}-{\bf
r_2}\mid$ (cf. equation 18 in FZH04).  Allowing the characteristic
size $R=R(z)$ and the mean neutral fraction $\bar\nf=\bar\nf(z)$ to
evolve with redshift introduces a new ambiguity in the redshift at
which $\bar\nf$ and $f$ is to be evaluated in $p_{\rm
same}=(1-\bar\nf) f$.  In practice, however, $\bar\nf$ and $R$ should
evolve only on long time--scales, comparable to the Hubble time.
Indeed, Barkana \& Loeb (2006) discuss an anisotropy in the
correlation function caused by this evolution, and the finite speed of
light.  As discussed in the Introduction, our focus here will be on
the anisotropies on smaller scales, comparable to the size of
individual quasar bubbles. The anisotropy from the cosmic evolution is
negligibly small on these scales (see Figure 3 in Barkana \& Loeb
2006), and we therefore ignore it here and use the simple expressions,
\begin{equation} 
 \xi_{xx}   = {\bar \nf^2 \over (1-p_{\rm same})},
\label{eq:corr_neu}
\end{equation}
and 
\begin{equation}
p_{\rm same} = (1-\bar \nf)f(r,R).
\label{eq:psame}
  \end{equation}
Note that the anisotropy we discuss below will arise entirely from the
dependence of $p_{\rm same}({\bf r_1,r_2})$ on ``orientation'' (i.e.,
the angle between ${\bf r_1-r_2}$ and the line of sight).

Let us consider various limits of the expressions in
equations~(\ref{eq:corr_neu}) and (\ref{eq:psame}): (i) for $r$
tending to zero $f$ tends to unity; (ii) as $r$ tends to infinity
$p_{\rm same}$ tends to zero 
and the correlation function  approaches  $\bar \nf^2$, as they
should; (iii) as the average  value of the neutral fraction tends to
zero, the correlation function vanishes, and (iv) and  when the
average value of the neutral fraction tends to unity, $\xi_{xx}$
should   approach  $\bar \nf^2$, as indeed it does.

\vspace{\baselineskip} 
\myputfigure{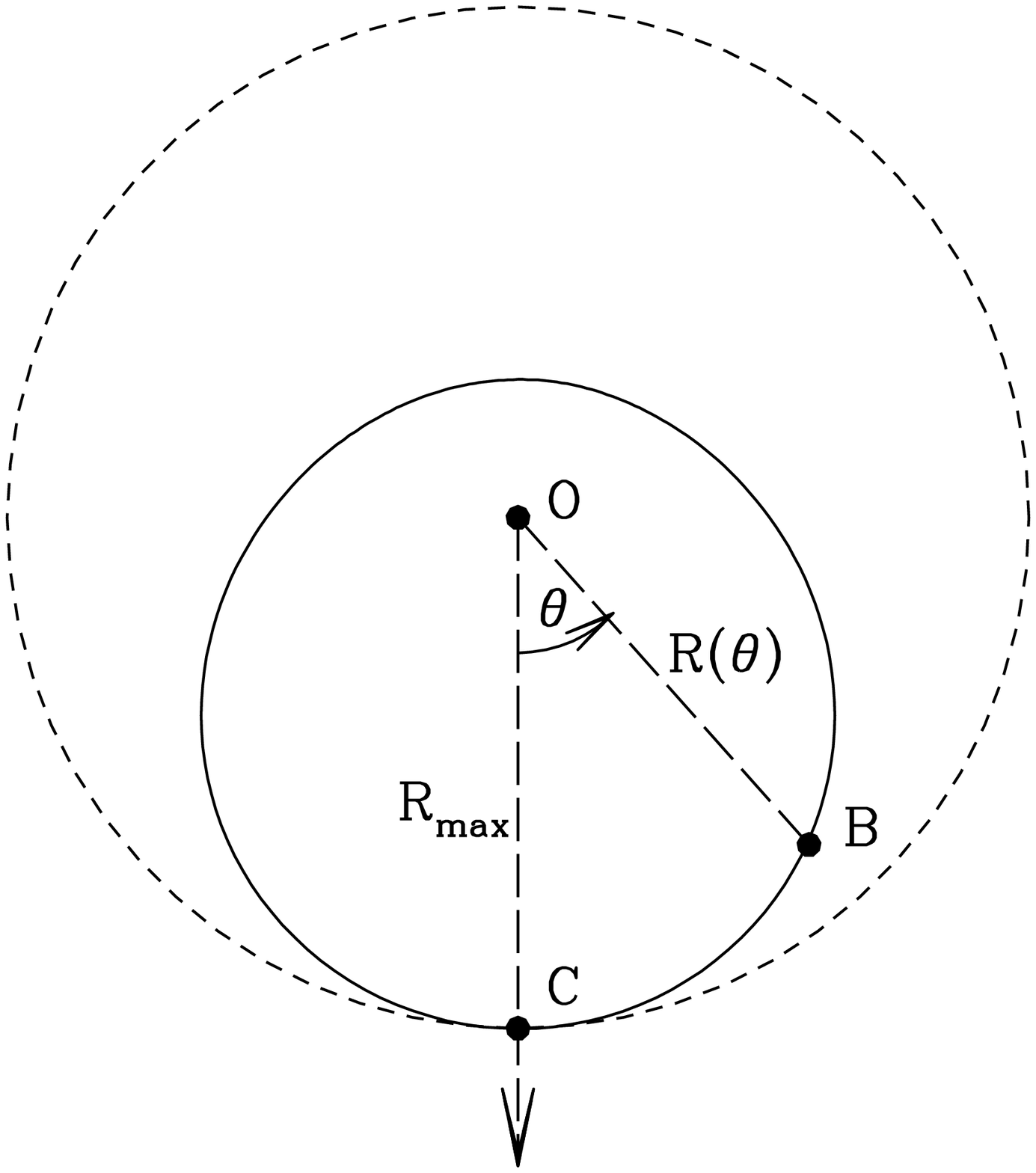}{3.2}{0.42}{-25}{-10}{0}
\vspace{\baselineskip} \figcaption{The apparent shape of the
Str\"omgren sphere. The solid curve shows the locus of points, in the
plane containing the ionizing source (O), from which light is observed
to arrive on Earth (toward the bottom of the plot) simultaneously.
The age of the Str\"omgren sphere is assumed to be $t_{\rm HII}=
10^7$yr, and the values of the other parameters are: clumping factor
$C=5$, neutral fraction $\bar \nf = 0.5$, source redshift $z = 10$ and
ionizing photon luminosity $S = 10^{57} \, \rm sec^{-1}$.  For
reference, the dashed curve shows a circle with radius $R_{\rm max}$
. The notation follows Yu (2005).
\label{fig:Rtheta0}}
\vspace{\baselineskip} 

\subsection{Light Travel Delay Anisotropy of Individual HII Regions}

Assuming that all ionizing sources carve out spherically symmetric
Str\"omgren spheres\footnote{Technically, these are not Str\"omgren
  spheres, since they likely do not reach equilibrium within the
  source lifetime, owing to the long recombination time in the IGM
  (e.g. Shapiro \& Giroux 1987); however, we follow here the
  wide--spread use of this misnomer.} in their rest frames (at least
statistically; we shall discuss this assumption in detail below), it
can be shown that as light from different points on the Str\"omgren
sphere is observed simultaneously by an observer on Earth, the sphere
will appear distorted.  The effect is, roughly, an elongation along
the line of sight; it is caused by the fact that photons observed at
increasingly large impact parameters away from the line of sight to
the source have traveled a longer path (Cen \& Haiman 2000; White et
al. 2003; Wyithe \& Loeb 2004c; Yu 2005). As a result, at large impact
parameters, we effectively observe the source at a younger age when it
had a smaller Str\"omgren radius.  

Estimating the apparent distortion entails computing the locus of all
points lying on the Str\"omgren sphere from which the light is
received at the same time. Expressions for this equal--arrival--time
surface, as a function of the angle relative to the line of sight have
been obtained in the case of a steady source (see Yu 2005 and
references therein).  We here adopt the results of Yu (2005), and
include a somewhat different, brief derivation for completeness.

Let us follow the notation of Yu (2005), and refer to the location of
the quasar (or any generic ionizing source) as ``$O$'' and to the
point on the Str\"omgren sphere that intersects the line of sight to
the quasar as ``$C$''. Let $\theta$ refer to the angle between $OC$
and $OB$, where $B$ is some other location on the Str\"omgren sphere's
surface (see Figure~\ref{fig:Rtheta0} for an illustration).  The
proper (not comoving) length of $OC$ is given by
\begin{equation}
R_{\rm max} = \int_{t_i}^{t_i + t_{\rm HII}} v(t') dt',
\end{equation}
where $v$ is the proper propagation velocity of the ionization front
in the rest frame of the quasar, $t_i$ is the cosmic time at which the
quasar switched on, and $t_i+t_{\rm HII}$ is the cosmic time at which
the quasar's photons reach the point $C$.  Here $t_{\rm HII}$ stands
for the age of the HII region at the time it is observed.  We note
that since photons take a finite time to travel between the source and
the edge of the HII region, this age is larger than the age $t_q$ of
the quasar, defined at the time it produced the photons that resulted
in the HII region: $t_{\rm HII}=t_q+R_{\rm max}/c$. Our aim is to find
$R(\theta)$, from the condition that photons traveling from the point
$B$ reach Earth simultaneously with photons from $C$.  This requires
that light leave the point $B$ earlier than $C$ by the difference in
light travel times,
\begin{equation}
\Delta t = {1 \over c}[R_{\rm max} - R(\theta) \cos(\theta)].
\end{equation} 
(Note that the angle subtended by the Str\"omgren sphere is small, and
we can safely assume that light rays reaching Earth from $C$ and $B$
are parallel.)  When the ionization front reaches the point $B$, the
age of quasar is smaller by this time delay, and $R(\theta)$ will be
smaller then $R_{\rm max}$; its value can be determined from the
implicit relation
\begin{equation}
R(\theta) = \int_{t_i}^{t_i + t_{\rm HII} - \Delta t} v(t') dt'.
\label{eq:Rtheta}
\end{equation}
with $R(0) = R_{\rm max}$.  

Given the velocity $v(t)$ of the ionization front, $R(\theta)$ can be
determined numerically from this relation (similar to the implicit
relation in eq.~3 of Yu 2005). A solution exists as long as $v <
c$. For $v = c$, a solution exists for $\theta \le \pi/2$ but not for
larger values of $\theta$ which just means we can only see the front
part of the Str\"omgren sphere.  The solution $R(\theta)$ tends to a
constant at late times, and $R(\theta)$ becomes independent of
$\theta$ as $v/c$ tends to zero (as should generally be the case at
late times).

\vspace{\baselineskip} 
\myputfigure{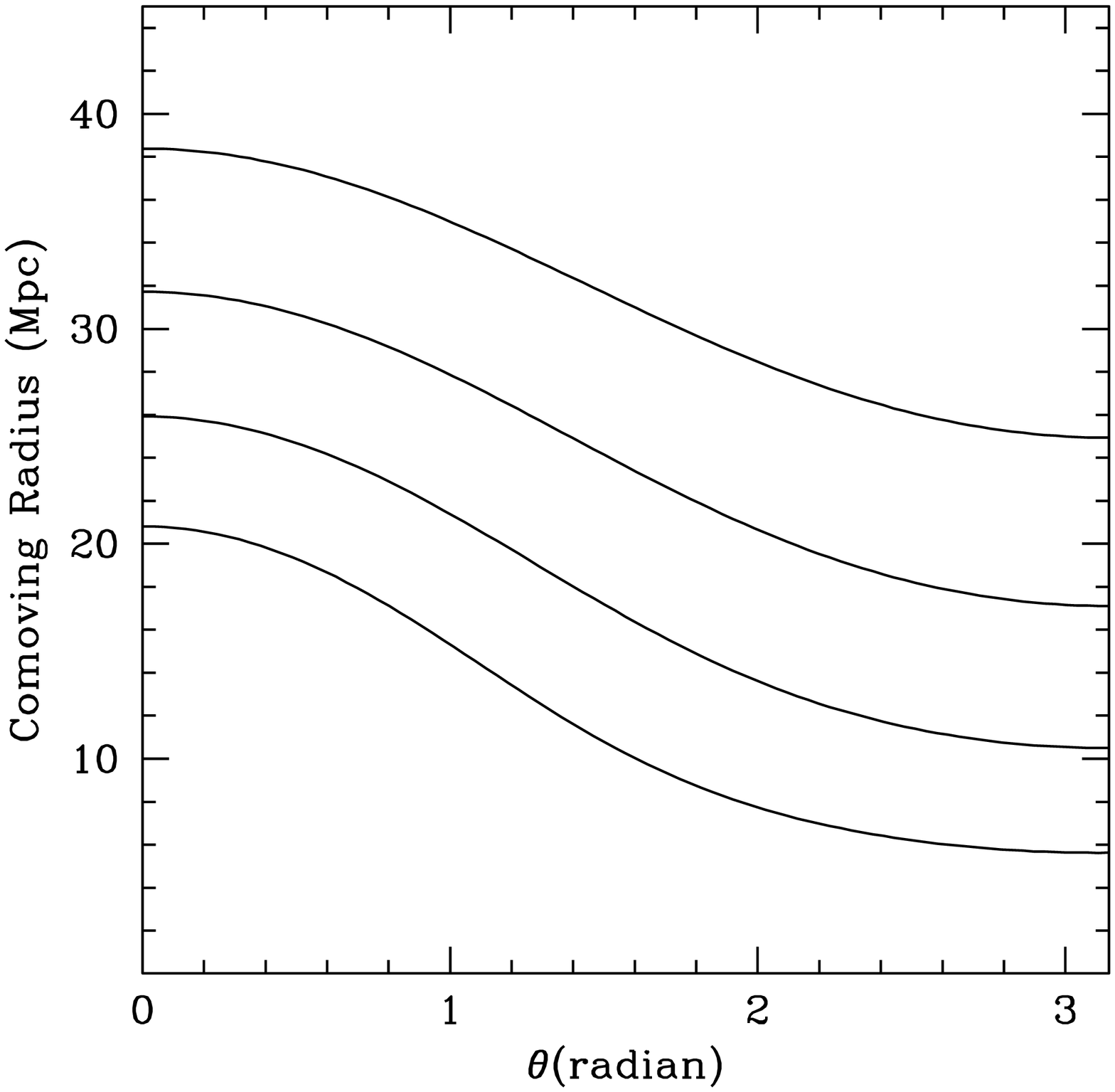}{3.2}{0.42}{-25}{-10}{0}
\vspace{\baselineskip} \figcaption{The comoving radius of the
Str\"omgren sphere $(1+z)R(\theta)$ is shown as a function of
$\theta$, the angle between the line of sight and the vector
connecting the position of the source with the point on the
Str\"omgren sphere.  The four curves show results for different ages
of the Str\"omgren sphere: from bottom to top the ages are $t_{\rm
HII}=\{1,1.5,2.25,3.75\} \times 10^7\, \rm years$.  The other
parameters are the same as in
Figure~\ref{fig:Rtheta0}. \label{fig:Rtheta}}
\vspace{\baselineskip}

The physical velocity of the growth of the Str\"omgren sphere $v =
dr/dt$ can be computed by (see e.g. White et al. 2003; Wyithe \& Loeb
2004b):
\begin{equation}
(S + 4\pi r^2 n_{\rm \scriptscriptstyle H} c)\left [{dr \over dt} -Hr \right ] = c\left(S - {4 \pi \over 3}r^3 n_{\rm \scriptscriptstyle H}^2 C \alpha_{\rm B} \right )
\label{strom_sp}
\end{equation}
Here $r$ is the physical (not comoving) radius of the Str\"omgren
sphere, $S$ is the ionizing photon luminosity of the central source,
$H=H(z)$ is the Hubble parameter at redshift $z$, $c$ is the speed of
light, $n_{\rm \scriptscriptstyle H}$ is the neutral hydrogen number
density, $\alpha_B$ is the case B recombination coefficient at the
temperature $10^4$K, and $C\equiv\langle n_{\rm \scriptscriptstyle
HII}^2\rangle / \langle n_{\rm \scriptscriptstyle HII}\rangle^2$.  is
the small-scale clumping factor of ionized gas inside the Str\"omgren
sphere.  This formula gives the correct limit of $dr/dt\rightarrow c$
as $r\rightarrow 0$, i.e. the Str\"omgren sphere expands at the speed
of light in the beginning.

In Figure~\ref{fig:Rtheta}, we show the comoving radius
$(1+z)R(\theta)$ for a set of fiducial parameters that could represent
a bright quasar at $z=10$, for four different ages of the HII region.
As the figure shows, the anisotropy can be large (of order unity) for
if the source is young ($t_{\rm HII}$=few$\times 10^7$yr).

As explained in the previous section, the effect of the
$\theta$--dependence of $R$ on the correlation function is that
$f(r,R)$ becomes a function of $\theta$.  Geometrically,
equation~(\ref{eq:frR}) represents the volume of overlap between two
spheres of radius $R$ that are a distance $r$ apart, divided by the
volume of a single sphere, $4\pi/3 R^3$.  This has to be replaced by
the new function $f(r,\theta)$, which represents the normalized
overlap volume between two objects whose shapes are described by
equation~(\ref{eq:Rtheta}).  Note that the shape has azimuthal
symmetry, but in general, the overlap volume does not. We numerically
compute this overlap volume, with the line connecting the ``centers''
of the two objects (i.e. the positions $O_1$ and $O_2$ of the sources)
of length $r$, and oriented at angle $\theta$ to the line of sight
(divided, again, by the volume of a single object).  The correlation
function (eq.~\ref{corr_fun_f}) can now be computed using
equations~(\ref{corr_fun_psi})--(\ref{eq:psame}), but replacing
$f(r,R)$ in equation~(\ref{eq:frR}) by the new $f(r,\theta)$.

\section{Results}
\label{sec:results}

The correlation function at a given redshift (eq.~\ref{corr_fun_f}),
including its anisotropy caused by the light travel delay, depends on
parameters the properties of the sources (such as their typical age,
luminosity, and number density) and of the IGM (such as its density,
mean neutral fraction, and clumping factor).  For a given reionization
history, the correlation function will be determined by averaging over
the distribution of the ages and luminosities of the sources that
co--exist in the IGM at any given redshift.

We first consider, in \S~\ref{subsec:toyresults}, simple toy models,
in which all ionizing sources (at a given redshift) are identical, and
propagate into a fully neutral IGM.  These results are intended to
illustrate the level of anisotropy expected from particular sources,
and the dependence of the anisotropy on various parameters.

In \S~\ref{subsec:realresults}, we present more realistic estimates of
the possible level of anisotropy.  These results differ from those in
the toy--model case, in that (i) we consider coexisting sources with a
range of luminosities, (ii) we include the dilution of the anisotropy
due to the presence of ``fossil'' HII regions, and (iii) we consider
the scenario in which the HII regions produced by bright quasars
expand into a medium that is already partially ionized by
pre--existing galaxies (and/or lower luminosity, non--relativistic
quasars).

Further complications will be discussed in \S~\ref{sec:discussion}
below.

\vspace{\baselineskip} 
\myputfigure{fig3.eps}{3.2}{0.32}{-25}{-10}{270}
\vspace{\baselineskip} \figcaption{The percentage change in the
correlation function, defined as $\Delta\xi/\xi\equiv
[\xi(0)-\xi(\theta)]/\xi(0) \times 100$, is plotted as a function of
$\theta$, the angle between the line of sight and the separation
vector ${\bf r_{12}}$ (note that this is different from the angle
subtended on the sky that is commonly used for angular correlation
functions).  Each curve corresponds to a different value of $r_{12}$
(comoving separation between two points).  The IGM is assumed to
contain randomly distributed, identical ionized bubbles.  The clumping
factor ($C=5$), neutral fraction ($\bar \nf = 0.5$), redshift ($z =
10$) and ionizing photon luminosity ($S = 10^{57} \, \rm sec^{-1}$)
are the same for each curve.  The four panels show the results for
different values of the age of the Str\"omgren sphere $t_{\rm HII}$;
clockwise from top left, the ages are $\{1,1.5,2.25,3.75\} \times
10^7\, \rm years$.  The value of $r_{12}$ and the normalization of the
temperature power spectrum, given by $C_(r_{12},\theta=0)$ in
equation~(\ref{corr_fun_f}), from bottom to top in each panel,
respectively, are: $r_{12} = \{5,10,15,20\} \, \rm Mpc$ and
$\{2.2\times 10^{-4}, 1.1\times 10^{-4}, 5.2\times 10^{-5}, 1.2\times
10^{-5}\} \, \rm K^2$;
$r_{12} = \{5,15,20,31\} \, \rm Mpc$ and $\{2.6\times 10^{-4},
6\times 10^{-5}, 5.2\times 10^{-5}, 4.4\times 10^{-6}\} \, \rm K^2 $; 
$r_{12} = \{12,25,35,45\} \, \rm Mpc$ and $\{1.9\times 10^{-4},
6.8\times 10^{-5}, 2.4\times 10^{-5}, 1.6\times 10^{-6}\} \, \rm K^2 $;
$r_{12} = \{35,50,58\} \, \rm Mpc$ and $\{5.6\times 10^{-5}, 1\times
10^{-5}, 1.8\times 10^{-6}\} \, \rm K^2 $. For reference, the dashed
curves in the top left panel show the anisotropy caused by
redshift-space distortion in $\xi_{\delta\delta}$
(eq.~\ref{corr_fun_rd}). The sign of the redshift-space distortion
anisotropy, as defined above, is everywhere negative, and the dashed
curves shows the absolute value of this change.  The length scale for
the curves, from bottom to top, is $\{5,15\} \, \rm Mpc$ and the
normalization of $\bar \nf^2\xi_{\delta\delta}(0)$ is $\{7\times
10^{-6},9 \times 10^{-7} \}\, \rm K^2$.  The anisotropies in
$\xi_{\delta\delta}$ and in $\xi_{xx}$ are comparable, but the
contribution of the density fluctuations $\xi_{\delta\delta}$ to
temperature anisotropy (eq. \ref{corr_fun_psi}) remains negligible for
the the range of length-scales shown in this figure.
  \label{fig:xi}}
\vspace{\baselineskip}

\subsection{Illustrative Toy Models}

\label{subsec:toyresults}

In this section, we consider a population of identical sources, in
order to isolate their contribution to the anisotropy in the more
realistic models described in the next subsection.  The most important
source parameters are the luminosity and lifetime; large anisotropies
are expected only for bright and short--lived sources, such as
luminous quasars. As we show below, the anisotropy from ionized
bubbles around galaxies or fainter quasars is expected to be
negligible.

\vspace{\baselineskip} 
\myputfigure{fig4.eps}{3.2}{0.32}{-25}{-10}{270}
\vspace{\baselineskip} \figcaption{All the parameters and the notation
are the same as Figure~\ref{fig:xi} except the photon ionizing
luminosity $S = 5 \times 10^{56} \, \rm sec^{-1}$.  The four panels
show the results for different values of the age of the Str\"omgren
sphere $t_{\rm HII}$; clockwise from top left, the ages are
$\{1,1.5,2.25,3.75\} \times 10^7\, \rm years$.  The value of $r_{12}$
and the normalization of the temperature power spectrum, given by
$C_(r_{12},\theta=0)$ in equation~(\ref{corr_fun_f}), from bottom to
top in each panel, respectively, are: $r_{12} = \{5,15,20\} \, \rm
Mpc$ and $\{2.1\times 10^{-4}, 4\times 10^{-5}, 5\times 10^{-5} \} \,
\rm K^2$;
$r_{12} = \{15,25,29\} \, \rm Mpc$ and $\{8\times 10^{-5},
1.2\times 10^{-5}, 5.2\times 10^{-6}\} \, \rm K^2 $; 
$r_{12} = \{25,31,38\} \, \rm Mpc$ and $\{5\times 10^{-5},
2\times 10^{-5}, 2.8\times 10^{-6}\} \, \rm K^2 $;
$r_{12} = \{31,38,48\} \, \rm Mpc$ and $\{5\times 10^{-5}, 2.4\times
10^{-5}, 3.2\times 10^{-6}\} \, \rm K^2 $.\label{fig:xi1}}
\vspace{\baselineskip}

\subsubsection{Quasar--like sources} 

Quasars can have high photon luminosities ($\simeq \hbox{0.5--5}
\times 10^{57} \, \rm sec^{-1}$) and short ages (of a few $\times 10^7
\, \rm years$; see, e.g., Haiman \& Cen 2002, or Martini 2004 for a
review).  In Figure~\ref{fig:xi} we plot the correlation function for
different choices of these parameters. The results from
Figure~\ref{fig:xi} can be briefly summarized as follows.

{\em (a) Level of Anisotropy.} The distortion from light travel delay
can lead to substantial anisotropy in the correlation function. The
level of anisotropy is $\sim 10\%$ or higher, if the source lifetimes
are $\lsim 2\times 10^7$ years, on scales $5$ Mpc$\lsim r_{12} \lsim
25$ Mpc.  The particular distortions in the shapes of the Str\"omgren
spheres translate to a characteristic $\theta$--dependence of the
anisotropy.

{\em (b) Dependence on Source Lifetime.} For fixed source luminosity,
and fixed separation $r_{12}$, the anisotropy generally diminishes as
the source lifetime gets longer. The sensitivity is very steep:
e.g. for $r_{12} \simeq 35 \,\rm Mpc$, the anisotropy for $t_{\rm HII}
= 3.75 \times 10^7 \, \rm years$ is about an order of magnitude
smaller than for $t_{\rm HII} = 1.5\times 10^7 \, \rm years$.  The
steep dependence of the anisotropy on the source lifetime is explained
by the fact that the HII regions initially expand rapidly at
relativistic speeds, but then become increasingly non--relativistic as
the source age increases. From equation~(\ref{strom_sp}), we can infer
that the transition occurs sooner for less luminous sources: the
critical radius is $r_{\rm nr} \simeq [S/(4\pi n_{\rm
\scriptscriptstyle HI} c)]^{1/2}$ and as $r \simeq ct$ during the
relativistic phase, the corresponding critical age of the source at
the transition scales as $t_{\rm nr} \propto S^{1/2}$. If sources have
typical life--times that are independent of $S$ (or at least scale
less steeply than $S^{1/2}$), then we can expect that bright sources
with $t\gsim t_{\rm nr}$ will be in their relativistic phase, whereas
fainter sources with $t\lsim t_{\rm nr}$ will have non--relativistic
and isotropic bubbles, with $t_{\rm nr}$ given by
\begin{equation}
t_{\rm nr}=4\times 10^7 {\rm yr}
\left(\frac{S}{2\times 10^{56}{\rm s^{-1}}}\right)^{1/2}
\left(\frac{1+z}{11}\right)^{-3/2}
\left(\frac{\nf}{0.5}\right)^{-1/2}.
\label{eq:Snr}
\end{equation}

{\em (c) Dependence on Scale.} The anisotropy shows a steady increase
with increasing $r_{12}$ and then peaks at $r_{12} \simeq 2 R_{c}$,
where $2R_{c}=R(0)+R(\pi)$ represents the diameter of a single HII
bubble.  For larger separations, the anisotropy decreases with
increasing $r_{12}$.

{\em (d) Dependence on Source Luminosity.} Figures~\ref{fig:xi} and
\ref{fig:xi1} show the change in anisotropy when the luminosity of the
quasars is changed by a factor of ten. A comparison of these figures
shows that a decrease in luminosity is generally associated with a
decrease in the anisotropy at a given length scale, for a given age of
the QSO. This is to be expected as the HII region of the QSO remains
relativistic for a longer period if the luminosity is increased
(eq.~\ref{eq:Snr}).

{\em (e) Dependence on IGM Clumping Factor.} We have repeated the
calculations in Figures~\ref{fig:xi} and \ref{fig:xi1} with the
clumping factor $C$ varied in the expected range of $1 \lsim C \lsim
20$.  The effect of an increased clumping factor is to slow down the
expansion of the ionization fronts (see eq.~\ref{strom_sp}), and
therefore to reduce the anisotropy. However, this effect is small
unless the ionization front speed is limited by recombinations, which
happens only at late times, and only if the clumping is near the
high--end of the expected range ($C\sim 20$).  Even in this case, we
found that the reduction in the level of anisotropy is at most $\sim
50\%$; the dependence on $C$ is negligibly small for $C\lsim 10$
(however, the clumping of the IGM has the additional, and more
significant effect of determining the ratio of fossil/active HII
regions; see discussion of this in \S~\ref{subsec:realresults} below).

{\em (e) Comparison to Redshift-Space Distortions.}  It is useful to
compare the anisotropy caused by quasars to the redshift-space
distortion induced anisotropy.  The two sources of anisotropies are,
in fact, coupled through the first term of
equation~(\ref{corr_fun_psi}). To disentangle these two effects, we
compute the correlation function in the density perturbations
$\xi_{\delta\delta}$ alone. The result is shown by the dashed curves
in the upper left panel in Figure~\ref{fig:xi}.  This panel shows that
the redshift-space distortion induced anisotropy in
$\xi_{\delta\delta}$ is significant, and at a comparable or larger
magnitude than the anisotropy in $\xi_{xx}$ caused by the quasars.
However, the anisotropy in the two-point correlation function is given
by the term $(1+\xi_{\delta\delta})\xi_{xx}$ in equation
(\ref{corr_fun_psi}). In the range of length scales shown in
Figures~\ref{fig:xi} and \ref{fig:xi1}, $\xi_{\delta\delta}\ll 1$, so
that the contribution of the density fluctuations $\xi_{\delta\delta}$
to the correlation function and its anisotropy remains negligible.  We
also note that for a range of scales, the sense of change of the
correlation function, as a function of the angle to the line of sight,
has the opposite sign compared with our signal from the finite light
travel time. The redshift-space distortion induced anisotropy is well
understood theoretically, and it would be interesting to estimate it
directly from the HI signal. However, the presence of even a small
amount of ionization-inhomogeneity induced anisotropy would suffice to
mask this anisotropy.

\subsubsection{Star--forming galaxies} 

In general, such sources are expected to be much less luminous than
bright quasars, and therefore equation~(\ref{eq:Snr}) suggests that
they can not cause any significant anisotropy.  We can estimate the
maximum photon luminosity of early galaxies by noting that $2\sigma$
peaks collapsing at $z=10$ correspond to halos with a total mass of
$\sim 10^8{\rm M_\odot}$, containing $\sim 10^7{\rm M_\odot}$ of
baryons.  If all baryons turn into stars with a Salpeter initial mass
function (IMF), each proton yields about $4000$ ionizing photons. This
efficiency can be further boosted by a factor of up to $\sim 20$, if
the stellar population consists entirely of massive, metal--free stars
(see Haiman \& Holder 2003 for further discussion and references).
Assuming the ionizing photons are produced within a single burst
lasting $10^7$ years, we find $S\approx 3\times 10^{54}{\rm s^{-1}}$.
From equation~(\ref{eq:Snr}), we conclude that even under these
extreme assumptions, the typical Str\"omgren spheres around these
sources will be expanding non--relativistically. We have explored a
range of luminosities $10^{53}\hbox{--}10^{54}{\rm s^{-1}}$ and ages
of $5 \times 10^6\hbox{--}5 \times 10^7$ years that could be expected
for the brightest star--forming galaxies, and a wide range of
parameters of the IGM, and indeed, we found that the anisotropy of the
correlation function is dominated by the redshift-space distortion.
Therefore if the star-forming galaxies are the exclusive agents of the
reionization process, the light travel time delay anisotropy is
unlikely to be detected.  Conversely, this implies that any secure
detection of the anisotropy would immediately reveal the presence of
bright, quasar--blown bubbles.

\subsection{Possible Anisotropy in More Realistic Scenarios}

\label{subsec:realresults}

As stated above, a more realistic model would have to include
averaging over a range of luminosities and ages at a given redshift.
The simple models shown in Figures~\ref{fig:xi} and \ref{fig:xi1},
however, already allow us to draw some basic conclusions.  First,
since the anisotropy decreases steeply with source lifetime, if all of
the ionizing sources have ages far exceeding a few $\times 10^7 \, \rm
years$, and/or they have ionizing luminosities significantly below a
few $\times 10^{56} {\rm sec^{-1}}$, then the correlation function
anisotropy will be negligible ($\ll 1\%$).  This critical value for
the lifetime coincides with expectations for bright quasars (see the
review by Martini 2004), whereas the critical luminosity is about an
order of magnitude below those of the already known SDSS quasars at
$z>6$.

Quasars of the required brightness almost certainly appeared only in
the late stages of reionization.  These relevant later parts of the
reionization history likely involved a mixture of both types of
sources discussed above. The process of reionization was likely
initiated by low--luminosity sources, such as star--forming galaxies
(with masses $\la 10^8 \, \rm M_\odot$), or smaller black holes (with
masses $\la 10^6 \, \rm M_\odot$; Madau et al. 2004; Ricotti \&
Ostriker 2004), at high redshift ($z\gsim 10$).  However, at somewhat
lower redshifts, but well before reionization is completed, bright
quasars could have formed in higher mass dark matter halos ($\simeq
10^{12} \, \rm M_\odot$) corresponding to the rarer high--$\sigma$
peaks of the underlying density distribution.  The unresolved soft
X--ray background puts a tight limit on the contribution of quasars to
reionization at $z\sim 6-20$, but does not rule out the above
scenario, in which quasars contribute a few ionizing photons per H
atom at $z\gsim 6$ (see Fig. 1 in Dijkstra, Haiman \& Loeb 2004).

The toy models above suggest that if bright quasars contribute a
non--negligible fraction of the ionizing photons produced at $z\gsim
6$, then the anisotropy of their Str\"omgren spheres could be
observable via the measurements of the 21cm correlation function. In
the rest of this section, we present somewhat more elaborate models,
to quantify the anisotropy in a scenario in which the universe is
partially ionized by faint sources (star-forming galaxies and/or
low--luminosity quasars) and the HII regions of bright, relativistic
QSOs, with a range of luminosities, are expanding into this partially
ionized medium.

More specifically, we make the following modifications to the toy
models of the previous section.

{\em (1) Pre--ionization by non-relativistic sources.} First, we
assume that at the epoch of interest, non--relativistic sources
(galaxies or low--luminosity quasars) had already ionized 60\% of the
IGM.  This is consistent with current constraints on the neutral
fraction around $z\sim 6$ (e.g. Fan, Carilli \& Keating 2006).  The
size of the HII regions of these bright QSOs is likely to be larger
than any other scale in the problem (e.g. compared to the clustering
scale of HII regions of other sources) and therefore one can assume
that at the scales comparable to the HII region of the bright QSO, the
only source of ionization inhomogeneity is owing to the presence of
the bright QSOs. Alvarez \& Abel (2007) and Lidz et~al. (2007) present
models which might bear out this picture (although, as these authors
argue, fluctuations from galaxies can be important in other contexts
involving smaller scales, such as interpreting Lyman $\alpha$
absorption spectra). For our purposes, the partial pre--ionization can
be regarded as uniform, and we therefore assume that its sole effect
is to reduce the HI density in the IGM (in eq.~\ref{strom_sp}).

{\em (2) Additional ionization by bright quasars.}  The ionization
fraction caused by bright QSOs could be a small fraction (say,
5--10\%) compared to the other sources (say, 60\%, as adopted above).
To be explicit, we set $\bar \nf = 0.05$ as our fiducial choice for
the quasar contribution, reduced by a factor of 10 from the
toy--models.  To motivate this choice, we note that each quasar with
the smallest luminosity of interest ($\gsim 5 \times 10^{56} \, \rm
sec^{-1}$) would contribute a newly ionized volume of $\sim 10^4 {\rm
Mpc^3}$. With the steepest allowed logarithmic slope for the quasar LF
at $z\sim 6$ (see discussion in \S~\ref{sec:discussion}), we expect a
space density of $\sim 10^{-5} {\rm Mpc^3}$ for these quasars.  The
choice of 5--10\% for the fraction of the volume ionized by these
relativistic sources is therefore about the maximum allowed by the
upper limits on their space density.  For reference, it is also useful
to note that at $z\sim 5-6$, the known population of bright and
detectable quasars contribute only about $1\%$ of the ionizing
background (e.g.  Madau, Haardt \& Rees 1999; Fan et al. 2001).
Sbrinovsky \& Wyithe (2007) used the observed quasar LF to explicitly
find an upper limit of 14.5\% on the contribution of luminous (but
below the detection threshold of SDSS) quasars at $z\approx 6$ to the
ionizing background.  In practical terms, this means that $p_{\rm
same}$ (eq.~\ref{eq:psame}) and the correlation function will now be
defined with respect to the ionization fraction caused by the bright
QSOs alone (e.g. 5\%).  The correlation function of the HI
fluctuations at the scale of the HII regions ($\gsim 10 \, \rm Mpc$)
of bright QSOs will be dominated by the ionization inhomogeneities
caused by these HII regions (the other component will come from the
density inhomogeneities which, as follows from Figure~\ref{fig:xi},
will contribute negligibly). This also means that the light--travel
time anisotropy at these scales will be mostly owing to the presence
of relativistic HII regions of the bright QSOs, even though they
contribute a small fraction to the ionization fraction.

{\em (3) A finite spread in quasar luminosities.}  It is obviously
unrealistic to assume that all quasars have the same luminosity.  To
relax this assumption, we extend our analysis to include HII regions
with a range of sizes. HII regions with volume $V$ then contribute an
ionized fraction $nV$, and one must compute the anisotropy by summing
over the contribution from different HII regions.  As shown in FZH04,
the existence of different bubble sizes can easily be incorporated, by
redefining quantities:
\begin{eqnarray}
x_e & =  & \sum_R x_e(R) \\
p_{\rm same}(r, \theta) &  =  &  \sum_R p_{\rm same}(r, R, \theta). \\
\end{eqnarray}
Here $R$ is the radius of the HII region corresponding to a particular
type of source, determined by the luminosity and lifetime of the
objects (note that this expression is strictly valid in the limit we
consider here, i.e. $nV\ll 1$).  For simplicity, we consider here only
a finite spread of luminosities, in the range $2 \times 10^{56} {\rm
  sec^{-1}} \lsim S \lsim 2 \times 10^{57} {\rm sec^{-1}}$.  Quasars
brighter than this range are already known to be rare. There could, of
course, be additional quasars below this luminosity range, but we
assume their HII regions would be nonrelativistic, and implicitly lump
any such quasars together with the galaxy pre--ionization discussed
above. We keep the lifetime fixed at $t_{\rm HII}=3\times 10^{7}$ yr
(in principle, one should also allow for a spread in ages between $0 <
t_{\rm HII} < 3.75\times 10^{7}$ yr, which would slightly increase the
anisotropy).  Within the $2 \times 10^{56} {\rm sec^{-1}} \lsim S
\lsim 2 \times 10^{57} {\rm sec^{-1}}$ range, we consider either a
flat or a steep luminosity function, $n(L)V(L) \propto L^0$ or
$n(K)V(L) \propto L^{-3}$.  The latter slope is motivated by current
empirical limits on the slope of the quasar LF at $z\sim 6$ (e.g. Fan,
Carilli \& Keating 2006), while the former (flat) slope is motivated
by the possibility that the high--$z$ quasar LF turns over and
flattens not too far below the SDSS detection threshold.

{\em (4) Fossil HII regions.} An important effect that will decrease
the anisotropy we predict for short--lived quasars is the presence of
``fossil'' HII regions around dead quasars.  At redshift $z=10$, for
our assumed $\Omega_bh^2$, the recombination time in the IGM is
$\approx 10^8 (C/5)^{-1}$ years.  This is longer by a factor $5
(t_{\rm HII}/2\times10^7{\rm yr})^{-1} (C/5)^{-1}$ than the ages of
the active HII regions.  The implication is that for $t_{\rm
  HII}=2\times10^7{\rm yr}$ and for $C=5$, at any given time, there
would be $\sim 5$ fossil HII regions, which have not yet fully
recombined, for every actively ionized Str\"omgren sphere.  Since
these fossil HII regions are not expanding, they will not produce an
apparent anisotropy, and they will dilute the predicted level of
anisotropy by a factor of $\sim 5$.  Several authors have attempted to
compute the gas clumping factor $C(z)$ from first principles, using
numerical simulations or semi--analytic arguments.  The overall
clumping factor at high $z$ is dominated by dense gas in collapsed
halos (Haiman et al. 2001) and the resulting small--scale clumping is
poorly understood.  Iliev, Scannapieco \& Shapiro (2005) find in a
simulation that the contribution from the low--density IGM to the
clumping at $z=10$ gives $C=8$; the dense gas inside halos can,
however, increase this value by a factor of a few, so that the
clumping factor may be as high as $C=20$ (Haiman et al. 2001; Iliev et
al. 2005).  On the other hand, the time taken by the fossil gas to
recombine will actually vary with the density, implying that the dense
parts of the fossil will recombine faster, while less dense parts will
linger longer.  Furthermore, the precise value of the relevant
clumping factor will also be modified due to clustering of gas near
quasars (one expects local overdensities, at least in the central
regions of the fossil), and will also depend on the instrumental
specifications.  The active/fossil bubble abundance ratio will also be
modified if the formation of quasars, within the last recombination
time, is skewed to more recent epochs.  In our fiducial model, we
assume $C=7$, which, for our choice of $t_{\rm HII}=3\times10^7{\rm
  yr}^{-1}$, implies that fossils outnumber active bubbles by a
factor of $\sim $three.  We include the presence of these fossils in
our calculations explicitly, treating them exactly as active HII
regions, except that we assume that they are isotropic (i.e. setting
$p_{\rm same}$ to be isotropic).  Note that the clumping factor, for
the purpose of computing the speed of the ionization front (in
eq.~\ref{strom_sp}) is different, and is still taken to be $C=10$.

\vspace{\baselineskip}
\myputfigure{fig5.eps}{3.2}{0.32}{-25}{-10}{270}
\vspace{\baselineskip} \figcaption{The expected level of anisotropy in
  our fiducial ``realistic'' model at redshift $z=7$. In this
  scenario, $60\%$ of the IGM volume is assumed to be already ionized
  by non--relativistic sources. Quasars with a range of ionizing
  luminosities ($2 \times 10^{56} {\rm sec^{-1}} \lsim S \lsim 2
  \times 10^{57} {\rm sec^{-1}}$) produce HII regions that ionize an
  additional 5\% of the IGM volume; isotropic fossil HII regions, left
  behind by dead quasars, are assumed to fill an additional 15\%.  The
  quasar lifetime is assumed to be fixed at $t_{\rm HII}=3\times
  10^{7}$ yr.  The clumping factor, for the purposes of the
  propagation of expanding ionization fronts, is taken to be $C=10$.
The scales in both panels, from bottom to top, are $r_{12} =
\{35,42,50\} \, \rm Mpc$.  Left panel: The curves assume a steep
quasar luminosity function, corresponding to $nV\propto L^{-3}$, and
the normalizations are $\{3\times 10^{-6}, 1.4\times 10^{-6},
3.7\times 10^{-7} \} \, \rm K^2$.
Right panel: The curves all correspond to a flat luminosity function,
i.e. a uniform distribution $nV\propto L^0$, and the normalizations
are $\{2.9\times 10^{-6}, 1.4\times 10^{-6}, 5\times 10^{-7} \} \, \rm
K^2$.
\label{fig:mixedres1}}
\vspace{\baselineskip}

The results of the calculations that take into account these
modifications are shown in Figure~\ref{fig:mixedres1}.  The
modifications (1)--(4) introduce several competing effects relative to
our toy models.  One the one hand, there is an increase in anisotropy
owing to the more rapid expansion of the bubbles in an already
pre-reionized medium.  On the other hand, there is a decrease in the
overall signal coming from these bubbles, as they contribute only a
small fraction of the total ionized fraction, and this contribution is
further diluted by the presence of fossil regions.  We find that the
effects of the finite spread of luminosities is less important,
although the steep LF within the luminosity range considered produces
a factor of $\approx$two lower anisotropy (because the fainter, less
relativistic sources will have a higher contribution). Overall, we
find that in the range of scales we considered, the level of the
anisotropy can still reach above $\sim$ 10 percent. The detectability
of this level of anisotropy will be discussed in the next section.

\section{Noise and Detectability}
\label{sec:noise}

The noise characteristics of interferometric experiments for the
detection of surface brightness fluctuations have been discussed in
detail in Fourier space in different contexts by various authors
(e.g. Bowman, Morales \& Hewitt 2006; Zaldarriaga et~al.  2004;
Bharadwaj \& Sethi 2001; White et~al. 1999; McQuinn et al. 2006).  The
anisotropy computation lends itself more readily to interpretation in
real space, we therefore estimate here the noise characteristics for
on-going and future interferometric experiments in real space.  The
surface brightness sensitivity $\Delta T_{\rm B}$ for detecting an
extended source (covering a solid angle equal to or larger than the
synthesized beam) is:
\begin{equation}
\Delta T_{\rm B}  = \Delta T_{\rm A} {\Omega_P \over \Omega_S}
\label{britem_sen}
\end{equation}
Here $\Omega_P$ and $\Omega_S$ are the solid angles of the primary and
the synthesized  beams, respectively. $ \Delta T_{\rm A}$, the antenna
temperature sensitivity is,
 \begin{equation}
\Delta T_{\rm A} \simeq {\sqrt{2} T_s \over \sqrt{\Delta \nu \Delta t
N^2}}
\end{equation}
Here $T_s$ is the system temperature, $\Delta \nu$ is the channel
width for the measurement, $\Delta t$ is the total integration time,
and $N$ is the total number of antennas.  As the noise computation is
valid for both filled antennas and dipole arrays, we refine our
definitions relevant to on-going and future experiments. The primary
beam for a filled--aperture experiment, such as GMRT, is the usual
``field of view''.  For an interferometer such as MWA, the primary
beam is the beam of each ``tile'' that contains sixteen dipoles, and
therefore $N$ is the total number of such tiles.  Another assumption
we shall make below is that the noise in each synthesized beam is
uncorrelated. This requires uniform coverage in the visibility
space. Both GMRT and MWA contain long baselines that can give a small
synthesized beam, but the noise in each pixel (synthesized beam) will
then be correlated. Therefore, in both cases, the the noise
computation here is applicable only to ``antennas'' within baselines
$\la 1 \, \rm km$.

The total number of independent (i.e. uncorrelated) pixels for a
frequency channel is $N_m \simeq \Omega_P/\Omega_S$. Assuming $N_\nu$
frequency channels, the total number of independent pixels in the data
cube is $N_t \simeq N_m N_\nu$. The quantity we wish to estimate is
the two point correlation function of the brightness temperature
fluctuations,
\begin{equation}
\xi_{12} = \langle \Delta T_B({\bf r_1})\Delta T_B({\bf r_2}) \rangle.
\label{eq:xi12}
\end{equation}
The observed temperature fluctuation $\Delta T_B({\bf r_1})$ contains
contributions from both signal and noise, but since noise is
uncorrelated between two pixels, for noise, $\xi_{12} = 0$, and
therefore the measured $\xi_{12}$ gives an unbiased estimator of the
signal.  We further assume that for the measured $\Delta T_B$ is
dominated by noise (or in other words, the signal cannot be directly
imaged in the given integration time; see discussion in \S~{sec:discussion}), 
and in our computations below, we will equate $\Delta T_B$ with the
noise. The quantity of interest is the variance of the two-point
function (eq. \ref{eq:xi12}).  This variance can be computed by using
the fact that each pair of measurements yields an uncorrelated random
variable. The variance in the average of $n$ such pairs is the
variance of each random variable, divided by the number of pairs (see
e.g. Papoulis 1984),
\begin{equation}
\delta  \xi_{12}^2 = {\sigma_\xi^2 \over n(r_{12},\theta)}
\end{equation}
Here $\sigma_\xi = \langle \Delta T_B^2 \rangle$ is the variance of a
single pair, and $n(r_{12},\theta)$ is the number of pairs for a fixed
$|{\bf r_{12}}|$.  Note that we explicitly write $n$ as a function of
$r_{12}$ and the angle $\theta$, to take into account all possible
correlations in the three--dimensional cube. The total number of pairs
for all the pixels in the data cube, $N_t$, is $N_{t,pair} =
N_t(N_t-1)/2$, which allows us to write $n(r_{12},\theta) = N_{t,pair}
\times f(r_{12},\theta)$, where $f(r_{12},\theta)$ is the fraction of
pairs within a given range of $r_{12}\pm \Delta r_{12}/2 $ and $\theta
\pm \Delta\theta/2$. This gives:
\begin{equation}
\delta  \xi_{12} = {2^{1/2} \Delta T_{A}^2 \over N_\nu f(r_{12},\theta)^{1/2}} \left ({\Omega_P \over \Omega_S} \right ). 
\label{nois_lev}
\end{equation}
The typical value of $f(r_{12},\theta)$ can be roughly estimated by
the following arguments. By definition, $f(r_{12},\theta)$ is the
fraction of pairs with a given $r_{12}$ and $\theta$. In practice, one
is likely to use, at least in the initial stages of an experiment with
small integration times, a small number of broad bins centered around
a set of values of $r_{12}$ and $\theta$, to estimate the signal.  For
example, one may choose to divide the range of lengths available with
a particular instrumental configuration into 10 bins with roughly
equal width $\Delta r_{12}$, and likewise divide the $\pi/2$ angular
range into 10 bins of equal $\Delta\theta$.  Using 100 such bands, a
typical value of $f(r_{12},\theta)$ will then be $\approx 10^{-2}$.
Of course, in reality, $f(r_{12},\theta)$ will depend on $r_{12}$ and
$\theta$, but this dependence can only be computed once the
instrumental configurations, and the choices of the bins, are
specified.

Here we list again the assumptions we have made to derive the above
expression. The main assumption is that the noise is uncorrelated
between pixels, or more specifically, the entire primary beam is
assumed to be filled with uncorrelated ``pixels'' of synthesized beam.
This, as mentioned above, requires uniform sampling in the visibility
plane with a small enough ``grid'' to sample the entire primary
beam. Whether this can be achieved depends on the array configuration.
As mentioned above, the on--going telescope GMRT, or the up--coming
MWA might be able to achieve such uniform sampling for baselines $\la
1\,\rm km$ (It should be noted that according to present MWA strategy,
the baseline distribution is not expected to be uniform, as we
implicitly assume for our estimates, but rather weighed towards
smaller baselines (roughly a $1/r^2$ distribution with a core of $10
\, \rm m$ (Bowman et~al.~2006).  McQuinn et~al. (2006) show that the
noise level expected for the weighed distribution are lower than for
the uniform distribution of baselines (Figure~6 of McQuinn
et~al. 2006).  Therefore our calculations here give a slight
overestimate of the expected noise levels.)  However, the statistical
homogeneity and isotropy of the reionization signal will allow one to
average over correlation function measurements from different primary
beams (i.e. over different patches of the sky).  As a result, even
when pixels in a single primary beam contain correlations, it is
possible to increase the number of uncorrelated correlation function
measurements, and therefore obtain the requisite number of
uncorrelated pixels.  On--going and future experiments, such as PAST
(Pen, Wu, Peterson 2004) and LOFAR (see {\tt www.lofar.org}), use a
large number of dipole antennas ($\simeq 10^4$), and our noise
estimate can be applied to these missions, as well.

We now give here numerical estimates of the noise, using
equation~(\ref{nois_lev}), for GMRT and MWA; similar estimates can be
made for missions such as PAST and LOFAR.  We assume the following
parameters common to GMRT and MWA: system temperature $T_s = 440 \,
\rm K$ (this system temperature corresponds to $z \simeq 8$ [Bowman
et~al. 2006]; we use this value throughout), total bandwidth, $N_\nu
\Delta \nu = 8 \, \rm MHz$ (note that the expected sensitivity depends
only on the total bandwidth and not on the channel width), total
observing time, $\Delta t = 10^6 \, \rm sec$, and $f(r_{12},\theta) =
10^{-2}$.  For GMRT, we adopt a primary beam of $\simeq 4^\circ$, a
synthesized beam of $\simeq 4'$ and assume $15$ antennas, which gives
$\Delta T_B \simeq 0.35 \, \rm K$ and $\delta \xi \simeq 2 \times
10^{-5} \, \rm K^2$.  For MWA, we assume a primary beam of $20^\circ$,
a synthesized beam of $4'$ and $500$ antennas (or tiles), which gives
$\Delta T_B \simeq 0.25 \, \rm K$ and $\delta \xi \simeq 2.5\times
10^{-7} \, \rm K2$ (Bowman et~al. 2006). 

Comparing these expected noise levels with the signal strength shown
in Figure~\ref{fig:mixedres1}, we conclude that while GMRT can not
detect the correlation function anisotropies, MWA has the capability
to do so.  For instance, the expected signal at $r_{12} \simeq \, \rm
35 Mpc$ is larger than the expected noise levels of MWA by a factor of
nearly eight.  The detection of the anisotropy at the level of
$2\hbox{--}10 \%$ will require noise levels smaller by a further
factor of nearly 10 to 50, which would require integration times
between $0.2\hbox{--}2 \times 10^7 \, \rm sec$.  These expectations
can be considerably improved by the using expected noise levels of
MWA5000 and SKA (Bowman et~al. 2006, McQuinn et~al. 2006).

One can compare the degree of difficulty involved in detecting the
anisotropy associated with finite-light-speed effects with the
anisotropy resulting from redshift-space distortion. As seen in
Figure~\ref{fig:xi}, even though the redshift-space anisotropy can be
large, it is generally associated with the sub-dominant component
(density fluctuations) of the overall signal. At scales $\simeq \, \rm
35 Mpc$, density perturbation-induced signal is more than order of
magnitude smaller than the signal in Figure~\ref{fig:mixedres1}. The
signal owing to density perturbations increases at smaller
scales. However, the forthcoming interferometers LOFAR and MWA are
better suited for detecting the signal above scales $\simeq 5 \, \rm
Mpc$ (see e.g. Figure~6 in McQuinn et~al. 2006); a future instrument
such as SKA, which is likely to have much better sensitivity at
smaller scales, is needed for detecting the redshift-space distortion
(McQuinn et~al. 2006). The anisotropy we discuss in this paper,
however, is dominant at large scales and, in particular, MWA is
ideally suited for detecting such a signal as it has much better
sensitivity at large scales, owing to its large primary beam.
 
The S/N estimate above asks whether the instrument can measure the
anisotropy on a particular scale $r_{12}$ and in a particular
direction $\theta$.  This estimate may be significantly more demanding
than a mere detection, or a crude characterization, of the anisotropy.
In reality, one can define much cruder measures of anisotropy to look
for -- say, dividing the angular range into a few ``quadrants'', and
combining the independent power spectrum measurements within each
quadrant, and for several different $r_{12}$. The effective number of
independent ($r,\theta$) combinations that one can combine will depend
on the actual observational strategy, and on the band power used, and
we leave more precise estimates to future work.
 
\section{Discussion}
\label{sec:discussion}

In deriving the anisotropy of the correlation function due to light
travel time delay, we assumed that the HII regions are spherically
symmetric around a given source of UV photons. For many reasons, the
HII regions will deviate from a spherical shape: (a) density
inhomogeneities around the source, (b) anisotropic emission from the
source, (c) mergers of HII regions.  We briefly discuss each of the
these in detail below.

{\it Density inhomogeneities}: From equation~(\ref{strom_sp}), it
follows that during the initial ultra-relativistic phase of expansion,
the velocity of the HII front is independent of density. As discussed
above, the HII front makes the transition from the relativistic to the
non-relativistic phase at the radius $r \propto n^{-1}_{\rm
\scriptscriptstyle HI}({\bf x})$. As a result, the HII region will
expand more slowly into higher density regions, and develops an
intrinsic anisotropy.  However, as the density field constitutes a
homogeneous and isotropic random process, averaging over a large
number of ionizing centers -- as will be required to measure the
correlation function -- will tend to cancel this effect.
Recombinations and radiative transfer effects will cause further
intrinsic anisotropies owing to density inhomogeneities (e.g., Bolton
\& Haehnelt 2007, Maselli et al. 2007), but these do not become
important until much later, well into the non--relativistic phase,
when the light travel time delay anisotropy is anyway negligible.

{\it Anisotropic emission}: Arguments similar to the previous case
apply equally to beamed or otherwise anisotropic emission, as well:
any intrinsic anisotropy in the source emission will be uncorrelated
with its orientation relative to the line-of-sight.  As a result,
averaging over a volume containing a large number of ionizing centers
will diminish this effect (producing no signal in the limiting case of
an very large survey volume).

{\it Anisotropy owing to bubble mergers}: Even if individual HII
regions are spherical, their mergers will result in strong
asymmetries, at least on scales comparable to the inter--distance
between the ionizing sources in a single bubble. The intrinsic shape
after the mergers of HII regions, or the level of its anisotropy, is
difficult to assess analytically, and would require numerical
simulations, which is beyond the scope of our work.  However, any
anisotropy caused by mergers also constitutes a homogeneous and
isotropic random process, uncorrelated with the line-of-sight. This
means that this anisotropy will be diluted as well, if measurements
average over many clusters of ionizing centers.

Mergers, however, will have some additional implications.  First, the
mergers will increase the characteristic bubble size, and therefore
also the length scale where the correlations (and their anisotropy)
peaks (e.g. FZH04).  Second, the light travel time delay anisotropy
will depend on the typical degree of synchronization between the
sources in a single super--bubble.  Predicting this synchronization
would require new assumptions about the source population, and
additional modeling, but the limiting cases are easily envisioned.  If
all sources in a bubble turned on simultaneously, this would be
similar to a single, but more luminous and anisotropic source, which
will boost the predicted signal (additional travel time delays between
the actual sources will again average out, if the sources are
isotropically distributed within the super--bubble).  On the other
hand, if the sources are turned on in a perfect ``sequence'', one
after the other, then the effect would be similar to increasing the
typical source lifetime (by a factor that equals the average number of
sources in a single bubble), without changing the luminosity.  This
will diminish the anisotropy signal, as discussed above. In
particular, if the source(s) in a bubble maintain a luminosity similar
to that of a single bright source for more than a $\sim 5\times 10^7$
years, the signal will become undetectable.  Whether this occurs would
be interesting to work out in specific models for bubble growth. The
situation should be possible to avoid when the filling factor of
ionized bubbles is small.  

\vspace{\baselineskip} 
\myputfigure{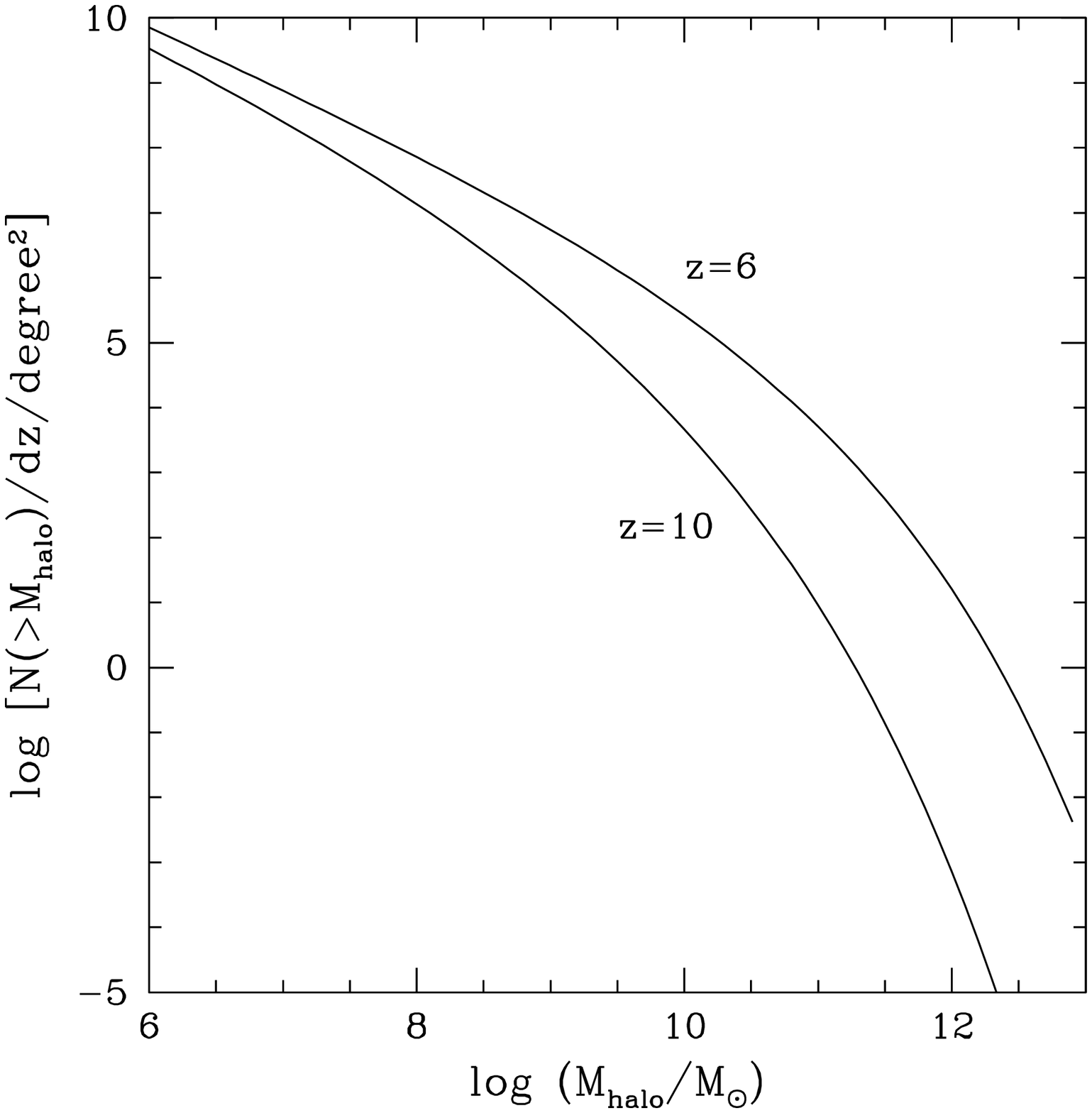}{3.2}{0.42}{-25}{-10}{0}
\vspace{\baselineskip} \figcaption{The number of dark matter halos
  per unit redshift and solid angle, at two different redshifts ($z=6$
  and $z=10$).
\label{fig:Nsources}}
\vspace{\baselineskip}

In summary, the anisotropy of the correlation function does not depend
crucially on the assumption of the sphericity of the HII region around
a source.  Detailed changes in the correlation function and its
anisotropy, owing to merging of sources are hard to assess but will
generally lead to an increase in correlation length. It can also
suppress the signal significantly if more than a few bright
quasar--like sources turn on, synchronized to within $\Delta t \approx
t_q$ (the lifetime of a single source), and cluster together in a
single bubble whose expansion is maintained for $\gsim 5\times 10^7$
years.  On the other hand, if the synchronization is either $\Delta t
\ll t_q$ or $\Delta t \gg t_q$, then we expect the anisotropy to
either increase, or remain unaffected, respectively.

In our models, we also implicitly assumed that the edges of the HII
regions are sharp.  In reality, the edges can be blurred if quasar
spectra are hard (Zaroubi \& Silk 2005; although see Kramer et
al. 2007 for a different conclusion) or if galaxies that are clustered
around the quasar contribute significant ionizing flux spread
spatially over an extended region (Wyithe \& Loeb 2006). In practice,
however, as long as the HII region boundaries are sharp enough to be
well defined, such blurring should not significantly affect our
conclusions. We also note that at least one of the $z>6$ quasars
appears to be surrounded by a sharp HII region, with a thickness
$\lsim 1$ (proper) Mpc for a radius of $\sim 5$ Mpc (Mesinger \&
Haiman 2004).  On the other hand, the random placement of the
instrumental resolution elements relative to the boundaries of the
quasar bubbles will effective further blur the edges of quasar
bubbles, and will also add shot noise to estimates of the anisotropy.

Whether the various intrinsic anisotropies can be averaged out depends
primarily in the number of bright sources within the survey volume.
Roughly, one expects that in order to detect a $\delta\xi/\xi = 10\%$
``systematic'' anisotropy, in the presence of random intrinsic
anisotropies of order unity, the survey has to contain at least $\gsim
(\delta\xi/\xi)^{-2}=100$ bubbles.  The number of bubbles is highly
uncertain, but we note that the lowest photon luminosity of interest
($\gsim 2\times10^{56}{\rm s^{-1}}$) corresponds to a luminosity that
is approximately a factor of $10-50$ below that of the bright quasars
detected in the SDSS (e.g. Wyithe \& Loeb 2004b).  The slope of the
quasar luminosity function at $z>6$ is expected to be steep, but
poorly constrained observationally.  Richards et al. (2006) placed an
upper limit on the slope at $4<z<5.4$ from the (lack of) gravitational
lensing, and found it to be flatter than $d\log n/d\log L\gsim
-4$. This limit, applied to $z\approx 6$, would allow the existence of
up to $\sim 100$ deg$^{-2}$ quasars that are sufficiently bright to
cause detectable anisotropy, or up to $3\times 10^4$ sources in one of
MWA's $300$deg$^{2}$ primary beam.

To provide another rough estimate of the possible number of
sufficiently bright sources, in Figure~\ref{fig:Nsources}, we show the
number of dark matter halos per unit redshift and solid angle, at two
different redshifts ($z=6$ and $z=10$), using the fitting formula from
Jenkins et al. (2001).  The abundance of the SDSS quasars implies that
they have host halos masses of $\sim 10^{12.8} {\rm M_\odot}$
(e.g. Haiman \& Loeb 2001).  Figure~\ref{fig:Nsources} shows that the
slope of the halo mass function at these high masses is $d\log N/d\log
M\sim -4$, similar (only slightly flatter) than the upper limit on the
slope of the quasar LF. Applying this scaling, we would expect
$\sim 10^{4.3}$ sources in the 300 deg$^{2}$ area of
one MWA primary beam. 
Because of the steepness of the halo mass function, however, this
estimate is very sensitive to the $M_{\rm bh}-M_{\rm halo}$ relation
at high redshift.  For example, if the quasar luminosity scales
linearly with BH mass $L_q\propto M_{\rm bh}$, but we had the steeper
scaling $M_{\rm bh}\propto M_{\rm halo}^{1.6}$ inferred for inactive
galaxies at lower redshifts (e.g. Ferrara 2002), the expected number
of sources would be reduced to $\sim 10^{2.3}$.  This would still
allow the detection of $\sim$few percent anisotropy, in the presence
of order unity variations in the shapes of individual bubbles, but
only marginally.

In our analysis above, we assumed, for simplicity, that the ionizing
sources are randomly distributed in space.  In reality, the sources
are likely to be located at the peaks of the density field and
therefore clustered.  As mentioned above, this will increase the
contribution of bright quasars to $\xi_{xx}$.  The clustering of the
sources, and the corresponding magnitude of the increase, could be
explicitly computed in a refined version of our model (see, e.g.,
equation 19 and related discussion in FZH04). Here we simply note that
the increase for rare sources should roughly trace their linear halo
bias (Sheth \& Tormen 1999) and we expect it to be about an order of
magnitude for sources residing in dark matter halos that correspond to
2--3$\sigma$ peaks (for explicit calculations of the impact of source
bias on the correlation function, see, e.g., Santos et al. 2003).
More interesting in the present context is that source clustering can
also modify the apparent anisotropy.  The three--dimensional spatial
correlation function of the sources themselves will appear isotropic
(neglecting peculiar velocities).  However, the probability that a
point is ionized is given by an integral of the space density of
ionizing sources, around the given point, over a volume whose shape is
anisotropic (given by eq.~\ref{eq:Rtheta}). The probability
distribution of sources within this volume will depend on both the
distance, and the angle with respect to the line of sight, from
another ionizing source.  This effect may not be negligible, since the
correlation length of the quasar distribution may approach the typical
sizes of ionized bubbles (quasars at $3.5<z<5.4$ already appear to
have a correlation length as large as 25 Mpc; Shen et al. 2007), and
will have to be included in future work and in analyzing actual 21cm
data.

As mentioned above, in our analysis, we neglected terms containing the
cross-correlation between density and the ionized fraction
(eq.~\ref{corr_fun_f}). Furlanetto et~al. (2005) argued that these
terms are likely to be small compared to the other two terms that we
retain, $\xi_{xx}\xi_{\delta\delta}$ and $(\xi_{xx} - \bar \nf ^2)$
(in eq.~\ref{corr_fun_psi}).  This is expected on the scale in
question ($\simeq 20 \, \rm Mpc$) as the leading terms we retain are
of order $0(\nf)$ or $0(\nf^2)$ depending on the value of $f(r,R)$.
The density-ionization fraction cross-correlation terms are of order
$0(\delta \nf)$, which is expected to be smaller than the terms
retained as $\delta \ll 1$ and $\delta/\nf \ll 1$ at scales of
interest.  The exact ratio of the cross-terms to the retained terms
can nevertheless only be reliably calculated with numerical
simulations.  In practice, within our model, $(\xi_{xx} - \bar \nf
^2)$ dominates over $\xi_{xx}\xi_{\delta\delta}$.  In their updated
bubble--growth model, McQuinn et al. (2005) explicitly compared
$(\xi_{xx} - \bar \nf ^2)$ and $\xi_{x\delta}$, over a range of
redshifts and length--scales.  They find that the cross-term is indeed
sub--dominant, although only by a factor of $\sim$ two at early stages
of reionization and on small scales (see, e.g., their Figure 2).  We
conclude that the cross--terms will not affect our estimates by more
than a factor of $\sim$two, but they will have to be included in a
more careful analysis of actual data.

In the context of the light travel time delay anisotropy, retaining
the density-ionized fraction cross-correlation can give rise to a new
anisotropy, owing to correlation between the redshift-space distortion
and the anisotropy due to light travel time delay. In linear theory,
the redshift-space distortion can readily be expanded into moments of
a Legendre transform, with only three of these moments non-vanishing
(see e.g. Hamilton 1998, eq.~(\ref{corr_fun_rd})). The light travel
time delay anisotropy can similarly be expanded into moments of
Legendre transform (this series will in general have a larger number
of non-vanishing moments) and the correlation with the redshift-space
distortion could be computed.

Finally, we note that the bright quasars producing the anisotropy in
the power spectrum should be directly detectable with a sensitive
future instrument, such as the {\it James Webb Space Telescope}
(e.g. Haiman \& Loeb 1998). Indeed, the 21cm maps may help identity
such quasars to begin with.  With the help of such identification, an
alternative method to search for the light travel time delay
anisotropy would be to stack the noisy 21cm tomographic images
centered around these quasars.  This method will require additional
modeling, in order to re--scale the sizes of the HII regions before
they are stacked.  Also, if the emission of quasars is anisotropic,
then this can introduce a selection effect: the optical selection will
preferentially detect those QSOs that appear brighter along the line
of sigh toward us.  This selection effect, if unaccounted for, will
mimic the effect of the anisotropy, since the transverse directions
around the quasar may see systematically lower fluxes.  We leave a
more detailed discussion of this stacking approach to future work.

\section{Conclusions}
\label{sec:conclude}

The time delay caused by finite light travel time across cosmological
HII regions distort their apparent shapes.  This effect may be
detectable in future redshifted 21cm observations for bright ionizing
sources during the various stages of reionization, and yield
constraints on the luminosity and ages of the sources, and the neutral
hydrogen density distribution in their surroundings (Wyithe \& Loeb
2004c; Yu 2005).  In principle, the distortion could be measured
directly in tomographic images of individual HII regions (Wyithe, Loeb
\& Barnes 2005).  Direct imaging of hundreds of sources at sufficient
S/N will, however, be unlikely to be achieved in forthcoming
experiments, such as LOFAR, PAST, or MWA, may have to await the
construction of SKA.

In this paper, we considered the detectability of this effect
statistically, through measuring the anisotropy in the three
dimensional 21cm power spectrum on a range of scales. We found that
the anisotropy is largest when HII regions expand at relativistic
speeds. Our results indicate that if bright quasars contributed
significantly (i.e. around 10 \% of the ionized fraction) to some
stage of reionization, then the finite-light-speed effect could be
observable in the anisotropy of the correlation function of the HI
distribution.  For quasar luminosities $\gsim 5 \times 10^{56} \, \rm
sec^{-1}$ and ages $\lsim 4 \times 10^7 \, \rm years$, we expect an
anisotropy of $\gsim 10 \%$ in the correlation function, as shown in
our results in Figures \ref{fig:xi} and \ref{fig:xi1}.  We also
compare this theoretical signal with the noise levels expected in
on-going and future radio interferometers that seek to detect this
signal. We show that on-going missions, such as MWA, might be able to
detect this effect.  A detection of this anisotropy would shed light
on the ionizing yield and age of the ionizing sources, and the
distribution and small--scale clumping of neutral intergalactic gas in
their vicinity.  In particular, a secure detection of this anisotropy
would immediately reveal the presence of a significant number of
bright quasars. These sources should be directly detectable with a
sensitive future instrument, such as the {\it James Webb Space
Telescope} (e.g. Haiman \& Loeb 1998), and indeed, the 21cm maps may
help identity such quasars to begin with.  With the help of such
identification, an alternative method to search for the light travel
time delay anisotropy would be to stack the noisy 21cm tomographic
images centered around these quasars.

\acknowledgments{We thank Rennan Barkana, Steve Furlanetto, Avi Loeb,
and Stuart Wyithe for useful comments on an earlier version of this
manuscript.  We also thank Matthew McQuinn and Miguel Morales for
useful discussions related to the sensitivity of MWA, and the referee
for comments that improved this paper.  ZH acknowledges partial
support by NASA through grant NNG04GI88G, by the NSF through grant
AST-0307291, and by the Hungarian Ministry of Education through a
Gy\"orgy B\'ek\'esy Fellowship.}

\end{document}